\documentclass[%
 reprint,
 amsmath,amssymb,
 aps,
]{revtex4-1}

\usepackage{graphicx}% Include figure files
\usepackage{dcolumn}% Align table columns on decimal point
\usepackage{bm}% bold math
\usepackage{epsfig}
\usepackage{float}
\usepackage{gensymb}
\usepackage{subfigure}
\usepackage{latexsym}
\usepackage{ragged2e}
\usepackage{textcomp}
\usepackage{gensymb}
\newcommand{\tbox}[1]{\mbox{\tiny #1}}
\newcommand{\bra}{\left\langle}
\newcommand{\ket}{\right\rangle}
\usepackage{etoolbox}
\usepackage{color}

\begin{document}

\title{Non-Hermitian diluted banded random matrices:\\
Scaling of eigenfunction and spectral properties}

\author{M. Hern\'andez-S\'anchez,$^1$ G. Tapia-Labra,$^1$ and J. A. M\'endez-Berm\'udez$^{1,2}$}
\address{Instituto de F\'isica, Benem\'erita Universidad Aut\'onoma de Puebla, Puebla 72570, Mexico \\
Escuela de F\'isica, Facultad de Ciencias, Universidad Nacional Aut\'onoma de Honduras, Honduras}

\date{\today}% It is always \today, today,

\begin{abstract}
Here we introduce the non-Hermitian diluted banded random matrix (nHdBRM) ensemble as the set of $N\times N$ 
real non-symmetric matrices whose entries are independent Gaussian random variables with zero mean and 
variance one if $|i-j|<b$ and zero otherwise, moreover off-diagonal matrix elements within the bandwidth $b$
are randomly set to zero such that the sparsity $\alpha$ is defined as the fraction of the $N(b-1)/2$ independent 
non-vanishing off-diagonal matrix elements.
By means of a detailed numerical study we demonstrate that the eigenfunction and spectral properties of the 
nHdBRM ensemble scale with the parameter $x=\gamma[(b\alpha)^2/N]^\delta$, where $\gamma,\delta\sim 1$.
Moreover, the normalized localization length $\beta$ of the eigenfunctions follows a simple scaling law: 
$\beta = x/(1 + x)$.
For comparison purposes, we also report eigenfunction and spectral properties of the Hermitian diluted banded 
random matrix ensemble.
\end{abstract}

\maketitle

\section{Introduction}

Random matrix (RM) models play a crucial role in the description of complex systems and complex 
processes~\cite{RMT}. 
Originating from the Gaussian Hermitian RM ensembles introduced by Wigner and Dyson~\cite{Metha}, 
RM models have served to reproduce the statistical properties of energy levels of diverse complex systems, 
such as heavy nuclei, quantized chaotic systems, disordered systems, and random networks~\cite{RMT}. 
In fact, RM models are not limited to describe the spectra of complex systems but also provide insights 
into a wide range of matrix-related quantities. In recent years, there has been a remarkable 
expansion of RM models to incorporate more sophisticated ensembles alongside the development of novel 
frameworks. Take, for instance, the problem of many-body localization~\cite{KKCA}, which exemplifies the 
advancement of RM models.

Even at the early years of RM modeling, Wigner himself foresaw the need for refinements to the original 
Gaussian ensembles, recognizing that they lacked certain properties exhibited by realistic physical systems. 
To address this, he proposed the so-called Wigner-banded RM model~\cite{W55,W57} (see also 
Refs.~\cite{RMT,WFL91,FLW91,FGIM93,CCGI93,CCGI96,F97,W00,W01}), incorporating elements like a 
bandwidth and an increasing diagonal. Indeed, the bandwidth, which measures the range of interactions, 
became the cornerstone for a variety of RM models emerging with distinct applications:
The power-law banded RM model~\cite{MFDQS96,EM08}, for instance, lies at the heart of simulating the 
Anderson metal-insulator transition, while the Banded Random Matrix (BRM) 
model~\cite{CMI90,EE90,FM91,CIM91,BMP91,FM92,MPK92,MF93,FM93,FM94,I95,MF96,CGM97,
KPI98,KIP99, P01,W02} serves to mimic the behavior of quasi-one-dimensional disordered wires.
The intricacies of many-body interactions in complex nuclei and many-body systems are also skillfully 
tackled through the embedded ensembles~\cite{MF75,BW03,K14}. Additionally, system-specific RM 
models~\cite{CK01,CH00}, tailored around banded Hamiltonian matrices corresponding to quantized 
chaotic systems, have also been proposed. 

These illustrations (see also 
Refs.~\cite{RMT,S94,FM95a,FM95b,S97,DPS02,KK02,S09,S10,CRBD16,MFRM16,PR93,FCIC96,MRV16,C17}), 
while far from exhaustive, indicate the wide variety of BRM models available to address a good number of 
different applications.

In addition, the analysis of diluted RM models has attracted significant interest as well, see for example 
Refs.~\cite{RB88,FM91b,FM91c,MF91,EE92,E92,JMR01,KSV04,K08,S09b,SC012,EKYY13,EKYY12,MAMRP15}.
In this context, we can mention the following models that include both sparsity and an effective bandwidth; 
i.e.~diluted BRM models: 
the Wigner-banded RM model with sparsity~\cite{PR93,FCIC96}, 
diluted power-law RM models~\cite{CRBD16,MRV16}, 
a diluted block-banded RM model~\cite{MFRM16}, and
the Hermitian diluted BRM ensemble~\cite{MFMR17b}.

It is important to stress that most of the studies mentioned above focus on Hermitian RM models even though
the Gaussian non-Hermitian RM ensembles were introduced already in the sixties by Ginibre~\cite{G65}.
This is relevant because non-Hermitian RM models may find direct applications in non-Hermitian 
physics, also known as non-Hermitian quantum mechanics (see e.g.~\cite{B07,M11}), which is a relatively 
new field of theoretical physics that challenges the 
conventional understanding of quantum mechanics by exploring the mathematical properties of non-Hermitian 
Hamiltonian operators. Indeed, non-Hermitian Hamiltonians can arise in systems such as open quantum systems, 
optical systems, and nonlinear systems, see e.g.~\cite{CGPR16,GMKMRC18}.

Remarkably, there exist several studies on diluted non-Hermitian RM ensembles, see 
e.g.~\cite{CES94,RC09,NM12,MNR19,BR19, AMRS20,PRRCM20,PM23,MMS24,MA24,DP21}, as well 
as a few papers on non-Hermitian banded RM models, see e.g.~\cite{DP21,J22}. However, we believe that 
more detailed studies of non-Hermitian banded RM models are still needed.

Consequently, with the aim of bridging the gap between diluted BRM models and non-Hermitian Hamiltonians, 
this paper explores the spectral and eigenfunction properties of a non-Hermitian diluted BRM ensemble
(see Sect.~\ref{scalingnHdBRM}), that we introduce here.
Moreover, for comparison purposes, we also review in detail the statistical properties of the Hermitian diluted BRM 
ensemble~\cite{MFMR17b} (see the Appendix~\ref{app}).
Our conclusions are drawn in Sect.~\ref{Conclusions}.

\section{Model definitions and Random Matrix Theory measures}

\subsection{Banded Random Matrix models}

The BRM ensemble is defined 
as the set of $N\times N$ real symmetric matrices whose entries are independent Gaussian 
random variables with zero mean and variance $1+\delta_{i,j}$ if $|i-j|<b$ and zero
otherwise. Hence, the bandwidth $b$ is the number of nonzero elements in the first matrix 
row which equals 1 for diagonal, 2 for tridiagonal, and $N$ for matrices of the Gaussian Orthogonal 
Ensemble (GOE)~\cite{Metha}. 
There are several numerical and theoretical studies available 
on this model, see for example 
Refs.~\cite{CMI90,EE90,FM91,CIM91,BMP91,FM92,MPK92,MF93,FM93,FM94,I95,MF96,CGM97,
KPI98,KIP99, P01,W02}. 
In particular, outstandingly, it has been found~\cite{CMI90,EE90,FM91,I95} 
that the eigenfunction properties of the BRM model, characterized by the {\it scaled localization 
length} $\beta$ (see Eq.~(\ref{betaS}) below), are {\it universal} for the fixed ratio 
\begin{equation}
X = \frac{b^2}{N} .
\label{XX}
\end{equation}
More specifically, it was numerically and theoretically shown that the scaling function
\begin{equation}
\beta = \frac{\Gamma X}{1+\Gamma X} ,
\label{scalingBRM}
\end{equation}
with $\Gamma\sim 1$, holds for the eigenfunctions of the BRM model, see also 
Refs.~\cite{FM92,MF93,FM93,FM94}. 
It is relevant 
to mention that scaling (\ref{scalingBRM}) was also shown to
be valid, when the scaling parameter $X$ is properly defined, for the kicked-rotator 
model~\cite{CGIS90,I90,I95} (a quantum-chaotic system characterized by a random-like 
banded Hamiltonian matrix), the one-dimensional Anderson model, and the Lloyd 
model~\cite{CGIFM92}.

On the other hand, 
the Hermitian diluted BRM (HdBRM) ensemble~\cite{MFMR17b} is defined by including sparsity 
into the BRM ensemble as follows: Starting with the BRM ensemble, off-diagonal matrix elements 
within the bandwidth $b$ are randomly set to zero such that the sparsity $\alpha$ is defined as the 
fraction of the $N(b-1)/2$ independent non-vanishing off-diagonal matrix elements. 
According to this definition, a diagonal random matrix is obtained
for $\alpha=0$, whereas the BRM model is recovered when $\alpha=1$.
Moreover, it was shown in Ref.~\cite{MFMR17b} that $\beta$ scales as
\begin{equation}
\beta = \frac{\gamma x^\delta}{1+\gamma x^\delta} .
\label{scalingHdBRM}
\end{equation}
In analogy with Eq.~(\ref{XX}), $x$ is defined as the ratio
\begin{equation}
x = \frac{b_{\mbox{\tiny eff}}^2}{N} ,
\label{x}
\end{equation}
where the {\it effective bandwidth}
\begin{equation}
b_{\mbox{\tiny eff}} \equiv \alpha b
\label{beff}
\end{equation}
is the average number of non-zero elements per
matrix row. In Eq.~(\ref{scalingHdBRM}), $\gamma,\delta\sim 1$.

Here, inspired by possible applications in non-Hermitian physics, we introduce the 
non-Hermitian diluted BRM (nHdBRM) ensemble as the non-Hermitian version of the HdBRM ensemble.
That is, the nHdBRM ensemble is the set of $N\times N$ real non-symmetric matrices 
whose entries are independent Gaussian random variables with zero mean and variance one if $|i-j|<b$ 
and zero otherwise, moreover off-diagonal matrix elements within the bandwidth $b$
are randomly set to zero such that the sparsity $\alpha$ is defined as the fraction of the 
$N(b-1)/2$ independent non-vanishing off-diagonal matrix elements.
According to this definition, a diagonal random matrix is obtained
for $\alpha=0$, whereas the real Ginibre ensemble (RGE)~\cite{G65} is recovered when $\alpha=1$ and $b=N$.
Notice that for $\alpha=1$ the nHdBRM ensemble becomes the non-Hermitian version of the BRM
ensemble that, as far as we know, has never been studied before.

Therefore, following the scaling studies of both the BRM 
ensemble~\cite{CMI90,EE90,CIM91,FM91,FM92,MF93,I95,CGM97,KIP99} and the HdBRM 
ensemble~\cite{MFMR17b}, here  we perform a detailed numerical study of eigenfunction and spectral 
properties of the nHdBRM ensemble. 
Moreover, since the study presented in Ref.~\cite{MFMR17b} included a limited number of random
matrix theory measures, here for completeness we also report eigenfunction and spectral 
properties of the HdBRM ensemble (see the Appendix~\ref{app}).

\subsection{Random Matrix Theory measures}

We use standard Random Matrix Theory (RMT) measures to characterize the eigenfunction and spectral 
properties of the matrices ${\bf nH}$ belonging to the nHdBRM ensemble.

Regarding eigenfunction properties, given the normalized eigenfunctions $\Psi^i$ 
(i.e.~$\sum_{m=1}^n\vert\Psi_m^i\vert^2 = 1$) of ${\bf nH}$, we compute 
the Shannon entropies~\cite{S48}
\begin{equation}
S_i = \sum_{m=1}^n\vert\Psi_m^i\vert^2 \ln\vert\Psi_m^i\vert^2
\label{shannon}
\end{equation}
and the inverse participation ratios~\cite{OH07}
\begin{equation}
\mbox{IPR}_i = \left[\sum_{m=1}^n\vert\Psi_m^i\vert^4\right]^{-1}.
\label{IPR}
\end{equation}
Both $S$ and $\mbox{IPR}$ measure the extension of eigenfunctions on a given basis. 

Particularly, Shannon entropies allows to compute the so-called entropic eigenfunction localization 
length, see e.g.~\cite{I90},
\begin{equation}
\label{lH}
\ell_N = N \exp\left[ -\left( S_{\tbox{RGE}} - \bra S \ket \right)\right] ,
\end{equation}
where $S_{\tbox{RGE}}\approx\ln(N/1.56)$~\cite{PRRCM20}, which is used here as a reference, 
is the average entropy of the eigenfunctions of the RGE. 
With this definition for $S$, when $\alpha=0$ or $b=1$, ${\bf nH}$ becomes a diagonal real random 
matrix (that is, a member of the Poisson Ensemble (PE)~\cite{Metha}) and the corresponding
eigenfunctions have only one non-vanishing component 
with magnitude equal to one; so $\bra S \ket=0$ and $\ell_N\sim 1$. 
On the contrary, 
when $\alpha=1$ and $b=N$ we recover the RGE and $\bra S \ket=S_{\tbox{RGE}}$; 
so, the {\it fully chaotic} eigenfunctions extend over the $N$ available basis states
and $\ell_N\approx N$.
Here, as well as in BRM model studies, we look for the scaling properties of the 
eigenfunctions of ${\bf nH}$ through the {\it scaled localization length}
\begin{equation}
\beta = \frac{\ell_N}{N} ,
\label{betaS}
\end{equation}
which can take values in the range $(0,1]$. 

Regarding spectral properties, given the complex spectrum $\{ \lambda_i \}$ ($i=1\ldots N$) of the 
non-Hermitian matrix ${\bf nH}$, we compute the ratio $r_{\mathbb{C}}$ between nearest- 
and next-to-nearest-neighbor eigenvalue distances, with the $i$--th ratio defined as~\cite{SRP20}
\begin{equation}
r_{\mathbb{C}}^i = \frac{\vert \lambda_i^{\mathrm{nn}}-\lambda_i\vert}{\vert \lambda_i^{\mathrm{nnn}}-\lambda_i\vert} ;
\label{rC}
\end{equation}
where $\lambda_i^{\mathrm{nn}}$ and $\lambda_i^{\mathrm{nnn}}$ are, respectively, the nearest and the 
next-to-nearest neighbors of $\lambda_i$ in $\mathbb{C}$.

Recently, the singular-value statistics (SVS) has been presented as a RMT tool able to properly 
characterize non-Hermitian RM ensembles~\cite{KXOS23}. So, we also use SVS here to characterize 
spectral properties of ${\bf nH}$ as follows: Given the ordered square roots of the real eigenvalues of 
the Hermitian matrix $({\bf nH})({\bf nH})^\dagger$, $s_1>s_2>\cdots >s_N$ (which are the singular 
values of ${\bf nH}$), we compute the ratio $r_{\tbox{SV}}$ between consecutive singular-value spacings, 
where the $i$--th ratio is given by~\cite{KXOS23}
\begin{equation}
r_{\tbox{SV}}^i = \frac{\mathrm{min}(s_{i+1}-s_i,s_{i}-s_{i-1})}{\mathrm{max}(s_{i+1}-s_i,s_i-s_{i-1})} .
\label{rSV}
\end{equation}
Above, as usual, $({\bf nH})^\dagger$ is the conjugate transpose of ${\bf nH}$.
Moreover, since for real matrices, as the ones we consider here, the conjugate transpose is just the 
transpose $({\bf nH})^\dagger=({\bf nH})^{\tbox{T}}$,
then, in what follows, the SVS concerns the spectra of $({\bf nH})({\bf nH})^{\tbox{T}}$.

\section{Scaling properties of non-Hermitian diluted banded random matrices}
\label{scalingnHdBRM}

In the following we use exact numerical diagonalization to obtain the eigenfunctions 
$\Psi^i$ ($i=1\ldots N$), the complex eigenvalues $\lambda_i$, and the singular values $s_i$  
of large ensembles of non-Hermitian matrices ${\bf nH}$ characterized by the parameter set ($N,b,\alpha$). 
For each of the averages reported below we used at least $5\times 10^5$ data values. 

\subsection{Scaling analysis of the localization length of eigenfunctions}

\begin{figure*}[ht]
\centering
\includegraphics[width=0.65\textwidth]{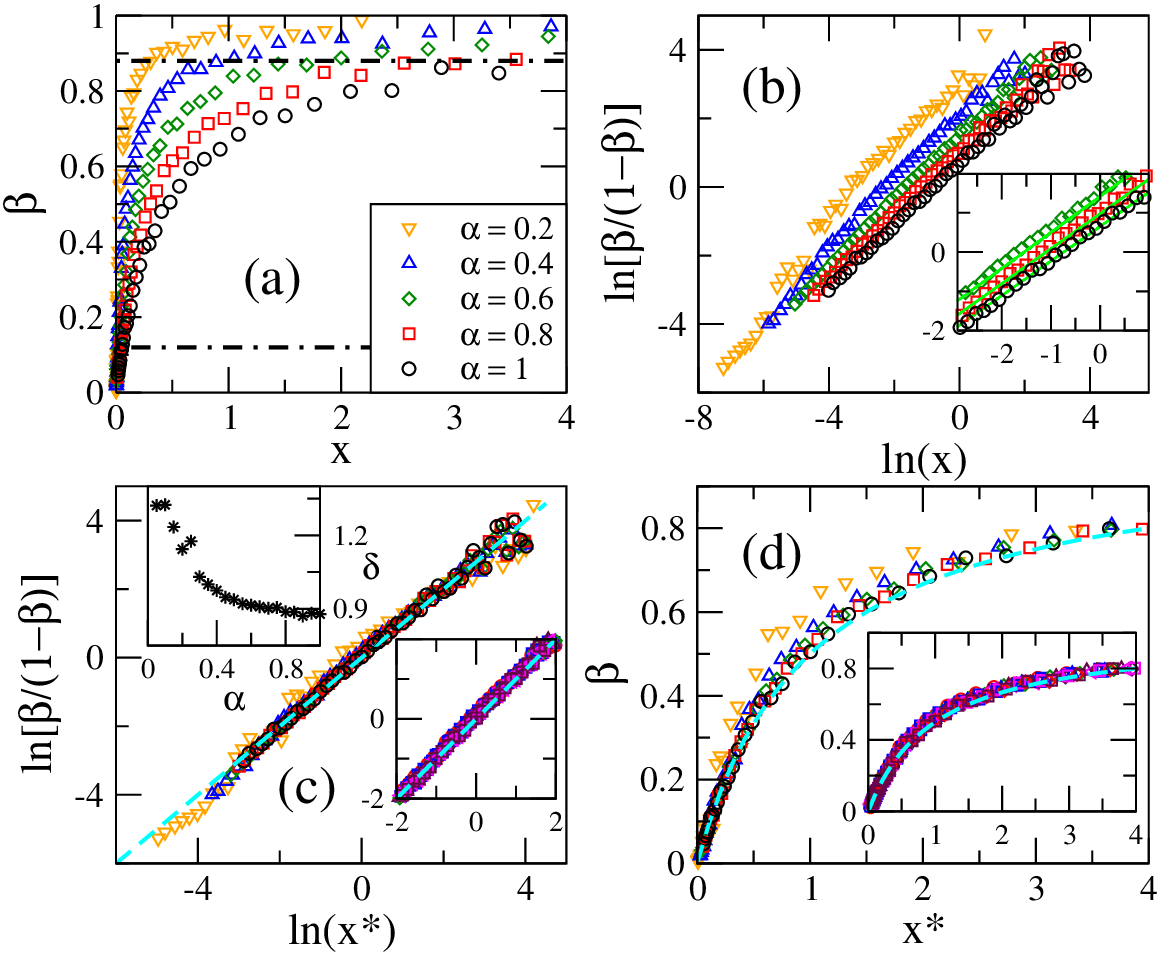}
\caption{
(a) Scaled localization length of eigenfunctions $\beta$ as a function of $x=b_{\mbox{\tiny eff}}^2/N$, 
$b_{\mbox{\tiny eff}}=\alpha b$, for the non-Hermitian diluted banded random matrix ensemble 
characterized by the sparsity $\alpha$. Several combinations of $(b,N)$ are used.
Horizontal dot-dashed lines at $\beta\approx 0.12$ and 0.88 are shown as a reference, see the text.
(b) Logarithm of $\beta/(1-\beta)$ as a function of $\ln(x)$. 
Inset: Enlargement in the range $\ln[\beta/(1-\beta)]=[-2,2]$ including data for 
$\alpha=0.6$, 0.8, and 1. Green-dashed lines are fittings of the data with Eq.~(\ref{betascaling2}).
(c) Logarithm of $\beta/(1-\beta)$ as a function of $\ln(x^*)$ [see Eq.~(\ref{betascaling3})]. 
Upper inset: Power $\delta$, from the fittings of the curves $\ln[\beta/(1-\beta)]$ vs.~$\ln(x)$ 
in the range $\ln[\beta/(1-\beta)]=[-2,2]$ with Eq.~(\ref{betascaling2}), as a function of $\alpha$. 
Lower Inset: Enlargement in the range $\ln[\beta/(1-\beta)]=[-2,2]$ including curves for 
$\alpha\in [0.5,1]$ in steps of $0.05$. Cyan-dashed lines in main panel and lower inset 
are Eq.~(\ref{betascaling3}).
(d) $\beta$ as a function of $x^*$. 
Inset: Data for $\alpha\in [0.5,1]$ in steps of $0.05$. Cyan-dashed lines in main 
panel and inset are Eq.~(\ref{betax*}).
}
\label{Fig01}
\end{figure*}
\begin{figure*}[ht]
\centering
\includegraphics[width=0.6\textwidth]{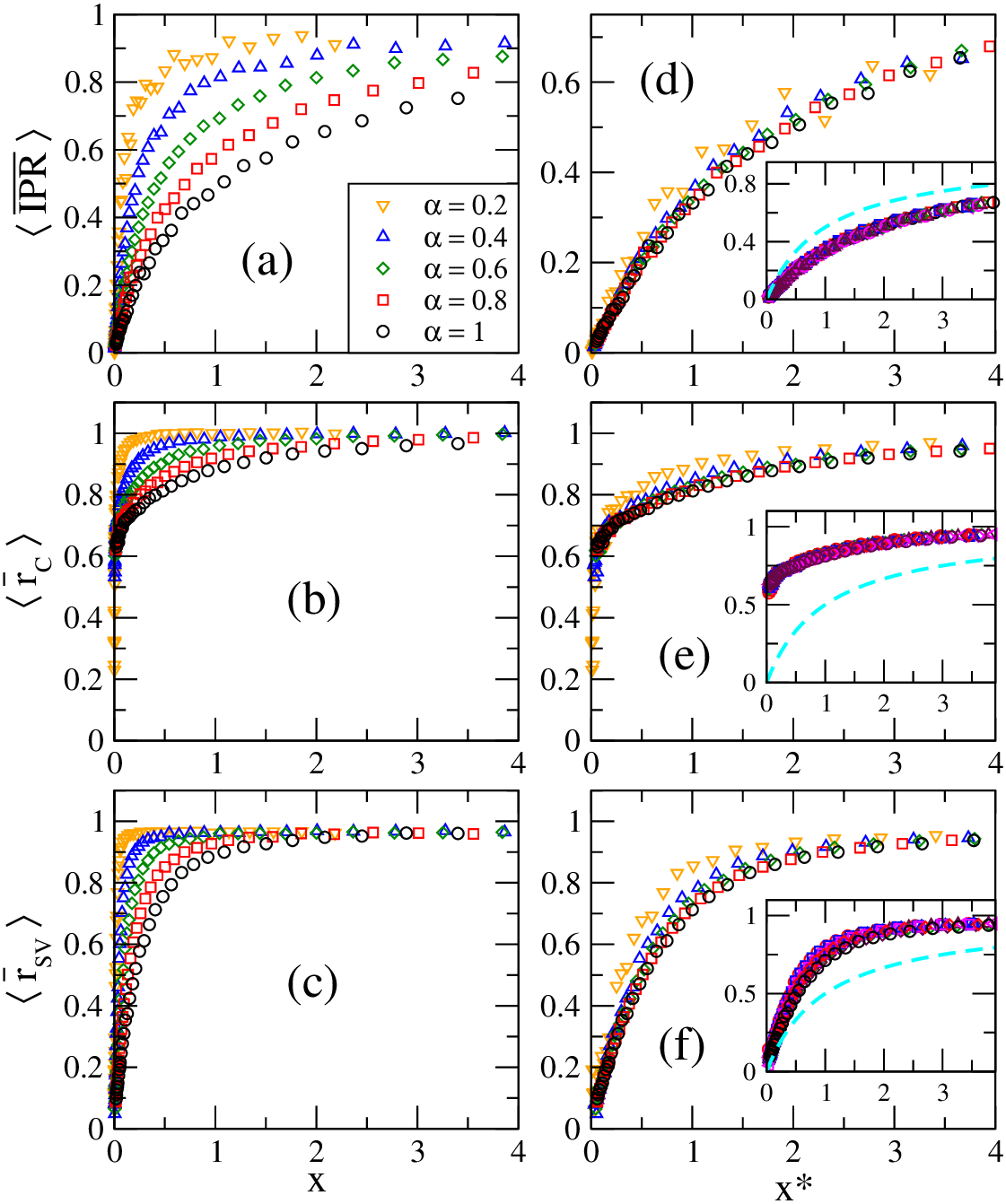}
\caption{
(a) $\left\langle \overline{\mbox{IPR}} \right\rangle$, 
(b) $\left\langle \overline{r}_\mathbb{C} \right\rangle$ and
(c) $\left\langle \overline{r}_{\tbox{SV}} \right\rangle$
as a function of $x$ for the non-Hermitian diluted banded random matrix ensemble 
characterized by the sparsity $\alpha$. 
(d) $\left\langle \overline{\mbox{IPR}} \right\rangle$, 
(e) $\left\langle \overline{r}_\mathbb{C} \right\rangle$ and
(f) $\left\langle \overline{r}_{\tbox{SV}} \right\rangle$
as a function of $x^*$.
Insets in (d-f): 
Data for $\alpha\in [0.5,1]$ in steps of $0.05$. Dashed lines are Eq.~(\ref{betax*}).
}
\label{Fig02}
\end{figure*}
\begin{figure*}[ht]
\centering
\includegraphics[width=0.65\textwidth]{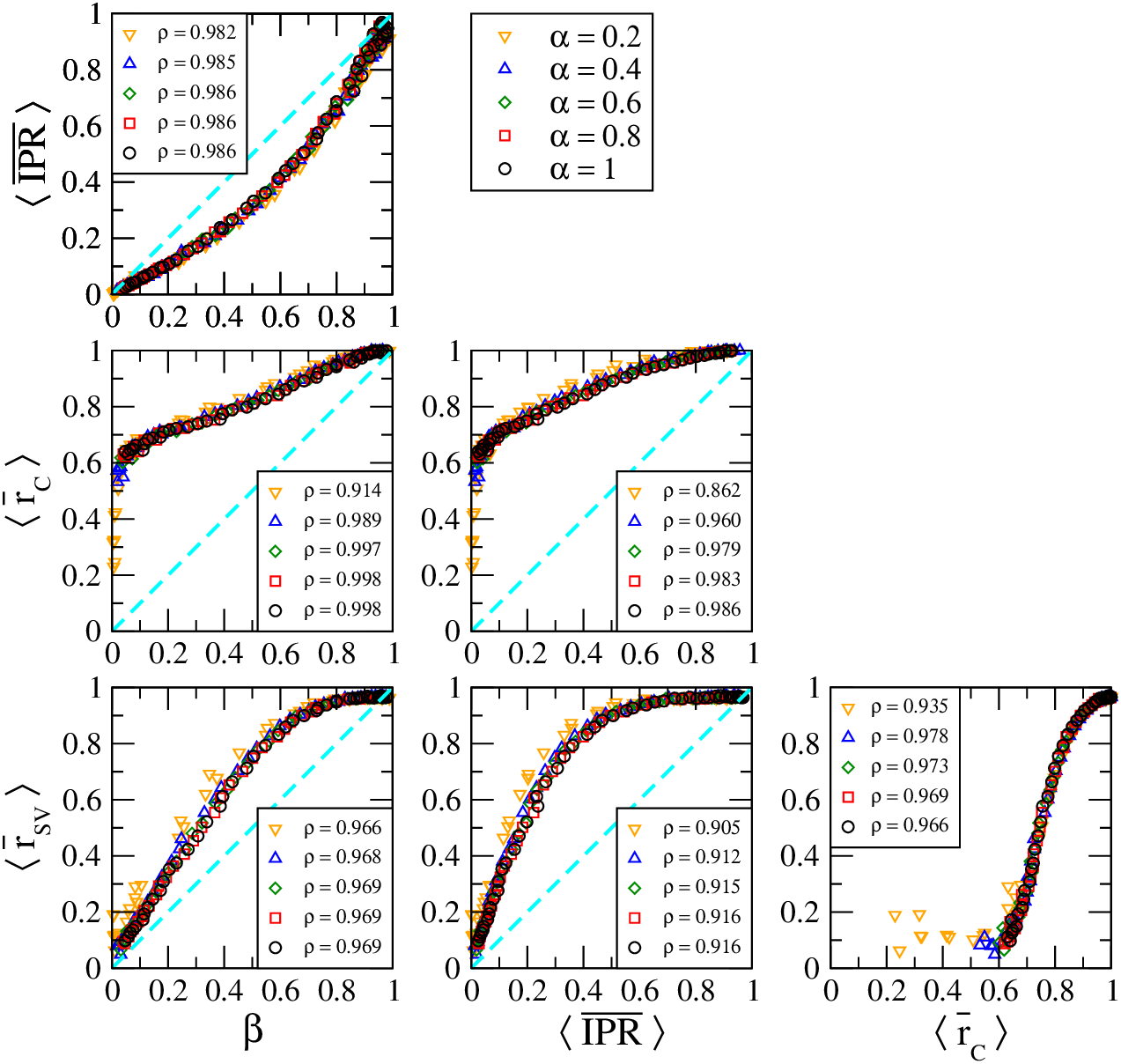}
\caption{
Scatter plots between $\beta$, $\left\langle \overline{\mbox{IPR}} \right\rangle$, 
$\left\langle \overline{r}_\mathbb{C} \right\rangle$ and
$\left\langle \overline{r}_{\tbox{SV}} \right\rangle$ for the non-Hermitian diluted banded 
random matrix ensemble. 
Pearson's correlation coefficients $\rho$ are reported in the corresponding panels.
}
\label{Fig03}
\end{figure*}

In Fig.~\ref{Fig01}(a) we plot $\beta$ as a function of $x=b_{\mbox{\tiny eff}}^2/N$, with
$b_{\mbox{\tiny eff}}=\alpha b$, for ensembles of matrices ${\bf nH}$ characterized by the sparsity 
$\alpha$. We observe that the curves of $\beta$ vs.~$x$ show similar functional forms however 
clearly affected by the sparsity $\alpha$: That is, for a fixed $x$, the smaller the value of $\alpha$ 
the larger the value of $\beta$. 
This panorama is equivalent to that reported for the HdBRM ensemble in Ref.~\cite{MFMR17b}.
Then, in Fig.~\ref{Fig01}(b) the logarithm of $\beta/(1-\beta)$ as a function of $\ln(x)$ is also 
presented. The quantity $\beta/(1-\beta)$ was useful in the study of the scaling properties of  
the BRM model~\cite{CMI90,FM92} because 
$\beta/(1-\beta) \propto x$, which is equivalent to scaling~(\ref{scalingBRM}), implies that a 
plot of $\ln[\beta/(1-\beta)]$  vs.~$\ln(x)$ is a straight line with unit slope. 
Even though, this statement was valid for the BRM model in a wide range of 
parameters (i.e.,~for $\ln[\beta/(1-\beta)]<2$) it does not apply to the nHdBRM ensemble; see
Fig.~\ref{Fig01}(b). In fact, from this figure we observe that plots of 
$\ln[\beta/(1-\beta)]$ vs.~$\ln(x)$ are straight lines (in a wide range of $x$) with 
a slope that depends (slightly but detectably) on the sparsity $\alpha$. 
Consequently, we put to test the scaling law 
\begin{equation}
\frac{\beta}{1-\beta} = \gamma x^\delta ,
\label{betascaling2}
\end{equation}
which is equivalent to scaling~(\ref{scalingHdBRM}), 
where both $\gamma$ and $\delta$ depend on $\alpha$. 

Indeed, Eq.~(\ref{betascaling2})
describes well our data, mainly in the range $\ln[\beta/(1-\beta)]=[-2,2]$, as can be 
seen in the inset of Fig.~\ref{Fig01}(b) where we show the data for 
$\alpha=0.6$, 0.8 and 1 and include fittings with Eq.~(\ref{betascaling2}).
We stress that the range $\ln[\beta/(1-\beta)]=[-2,2]$ corresponds to a
reasonable large range of $\beta$ values, $\beta\approx[0.12,0.88]$, whose bounds
are indicated with horizontal dot-dashed lines in Fig.~\ref{Fig01}(a).
Also, we notice that the power $\delta$, obtained from the fittings of the data 
using Eq.~(\ref{betascaling2}), is quite close to unity for all the sparsity
values we consider here; see the upper inset of Fig.~\ref{Fig01}(c).

Therefore, from the analysis of the data in Figs.~\ref{Fig01}(a,b), we can write down 
a {\it universal scaling function} for the scaled localization length $\beta$ of the
nHdBRM model as
\begin{equation}
\frac{\beta}{1-\beta} = x^* \ , \qquad x^*\equiv \gamma x^\delta .
\label{betascaling3}
\end{equation}
To validate Eq.~(\ref{betascaling3}) in Fig.~\ref{Fig01}(c) we present again the
data for $\ln[\beta/(1-\beta)]$ shown in Fig.~\ref{Fig01}(b) but now as a function 
of $\ln(x^*)$. We do observe that curves for different values of $\alpha$ fall on 
top of Eq.~(\ref{betascaling3}) for a wide range of the variable $x^*$.
Moreover, the collapse of the numerical data on top of Eq.~(\ref{betascaling3}) 
is excellent in the range $\ln[\beta/(1-\beta)]=[-2,2]$ for $\alpha\ge 0.5$, as
shown in the lower inset of Fig.~\ref{Fig01}(c).

Finally, we rewrite Eq.~(\ref{betascaling3}) into the equivalent, but explicit, 
scaling function for $\beta$:
\begin{equation}
\beta = \frac{x^*}{1+x^*} .
\label{betax*}
\end{equation}
In Fig.~\ref{Fig01}(d) we confirm the validity of Eq.~(\ref{betax*}).
We would like to emphasize that the universal scaling given in Eq.~(\ref{betax*})
extends outsize the range $\beta\approx[0.12,0.88]$, for which Eq.~(\ref{betascaling2})
was shown to be valid, see the main panel of Fig.~\ref{Fig01}(d). Furthermore, 
the collapse of the numerical data on top of Eq.~(\ref{betax*}) is remarkably good for 
$\alpha\ge 0.5$, as shown in the inset of Fig.~\ref{Fig01}(d).

\subsection{Other RMT measures}

Now we complete the analysis of eigenfunction and spectral properties of the nHdBRM
ensemble by computing 
the average inverse participation ratios $\left\langle \mbox{IPR} \right\rangle$ as well
as the average ratios $\left\langle r_\mathbb{C} \right\rangle$ and 
$\left\langle r_{\tbox{SV}} \right\rangle$, see Eqs.~(\ref{IPR},\ref{rC},\ref{rSV}).
Moreover, we conveniently normalize these averages as follows:
\begin{equation}
\langle \overline{\mathrm{IPR}}\rangle= \frac{\langle \mathrm{IPR} \rangle-\mathrm{IPR}_{\mathrm{PE}}}{\mathrm{IPR}_{\mathrm{RGE}}-\mathrm{IPR}_{\mathrm{PE}}} ,
\label{anIPR}
\end{equation}
\begin{equation}
\langle \overline{r}_\mathbb{C}\rangle= \frac{\langle r_\mathbb{C}  \rangle-r_{\mathbb{C}_{\mathrm{PE}}}}{r_{\mathbb{C}_{\mathrm{RGE}}}-r_{\mathbb{C}_{\mathrm{PE}}}} 
\label{anrC}
\end{equation}
and
\begin{equation}
\langle \overline{r}_{\tbox{SV}}\rangle= \frac{\langle r_{\tbox{SV}}  \rangle-r_{{\tbox{SV}}_{\mathrm{PE}}}}{r_{{\tbox{SV}}_{\mathrm{RGE}}}-r_{{\tbox{SV}}_{\mathrm{PE}}}} ,
\label{anrSV}
\end{equation}
such that they all take values in the interval $[0,1]$, so they can be directly compared with $\beta$.
The reference values used in Eqs.~(\ref{anIPR}-\ref{anrSV}), 
corresponding to the PE and the RGE, are reported in Table~\ref{T1}.

\begin{table}[b!]
\caption{
Reference average values for the Poisson ensemble and the real Ginibre ensemble,
used in Eqs.~(\ref{anIPR}-\ref{anrSV}), and the Gaussian orthogonal ensemble, to be used in 
the Appendix~\ref{app}.
}
\label{T1}
\begin{tabular}{ c | c | c | c | c  }  
\hline
 & $\mathrm{IPR}$  & $r_\mathbb{C}$ & $r_{\tbox{SV}}$ & $r_\mathbb{R}$ \\
\hline
PE    & 1 & 0.5~\cite{PRRCM20} & 0.5~\cite{MA24} & 0.386~\cite{ABGR13} \\
RGE & N/2.04~\cite{PM23} & 0.737~\cite{PRRCM20} & 0.569~\cite{MA24}  & -- \\
GOE & N/3~\cite{OH07} & 0.569~\cite{PRRCM20} & -- & 0.536~\cite{ABGR13} \\
\hline
\end{tabular}
\end{table}

Then, in Figs.~\ref{Fig02}(a-c) we plot the normalized measures 
$\left\langle \overline{\mbox{IPR}} \right\rangle$, 
$\left\langle \overline{r}_\mathbb{C} \right\rangle$ and
$\left\langle \overline{r}_{\tbox{SV}} \right\rangle$, respectively,
as a function of $x$ for the nHdBRM ensemble characterized by the sparsity $\alpha$.
The panorama shown in Figs.~\ref{Fig02}(a-c) for $\left\langle \overline{\mbox{IPR}} \right\rangle$, 
$\left\langle \overline{r}_\mathbb{C} \right\rangle$ and
$\left\langle \overline{r}_{\tbox{SV}} \right\rangle$ is equivalent to that observed for
$\beta$ in Fig.~\ref{Fig01}(a): 
The curves of $\left\langle \overline{X} \right\rangle$ vs.~$x$ show similar functional forms however 
clearly affected by the sparsity $\alpha$. 
Here, $X$ represents $\mbox{IPR}$, $r_\mathbb{C}$ and $r_{\tbox{SV}}$.
Also, for a fixed $x$, the smaller the value of $\alpha$ the larger the value of $\left\langle \overline{X} \right\rangle$. 
This observations allows us to surmise that the scaling parameter of $\beta$, $x^*$, may also
serve as scaling parameter of $\left\langle \overline{X} \right\rangle$.
To verify this assumption, in Figs.~\ref{Fig02}(d-f) we plot again  
$\left\langle \overline{\mbox{IPR}} \right\rangle$, 
$\left\langle \overline{r}_\mathbb{C} \right\rangle$ and
$\left\langle \overline{r}_{\tbox{SV}} \right\rangle$, respectively,
but now as a function of $x^*$.
Indeed, since the curves $\left\langle \overline{X} \right\rangle$ vs.~$x^*$ fall one on top of the other 
mainly for $\alpha\ge 0.5$, see the corresponding insets, we conclude that $x^*$ scales 
$\left\langle \overline{\mbox{IPR}} \right\rangle$, 
$\left\langle \overline{r}_\mathbb{C} \right\rangle$ and
$\left\langle \overline{r}_{\tbox{SV}} \right\rangle$
as good as it scales $\beta$.
From Fig.~\ref{Fig02} we also observe that the curves
$\left\langle \overline{r}_\mathbb{C} \right\rangle$ vs.~$x^*$ and 
$\left\langle \overline{r}_{\tbox{SV}} \right\rangle$ vs.~$x^*$
are above Eq.~(\ref{betax*}), which is included as dashed lines.
This also means that the spectral properties of the nHdBRM model approach the RGE limit faster 
than the eigenfunction properties.

From Figs.~\ref{Fig01} and~\ref{Fig02} we can also see that all quantities ($\beta$, 
$\left\langle \overline{\mbox{IPR}} \right\rangle$, 
$\left\langle \overline{r}_\mathbb{C} \right\rangle$ and
$\left\langle \overline{r}_{\tbox{SV}} \right\rangle$)
appear to be highly correlated. Therefore, in Fig.~\ref{Fig03} we present
scatter plots between $\beta$, $\left\langle \overline{\mbox{IPR}} \right\rangle$, 
$\left\langle \overline{r}_\mathbb{C} \right\rangle$ and
$\left\langle \overline{r}_{\tbox{SV}} \right\rangle$ for the nHdBRM ensemble for
several values of $\alpha$, where the high correlation between them is evident.
To quantify the correlation among these quantities, in the panels of Fig.~\ref{Fig03} we report the
corresponding Pearson's correlation coefficient $\rho$, which turns
out to be relatively large in all cases; i.e.~$\rho>0.9$.

\section{Discussion and conclusions}
\label{Conclusions}

In this work, by using extensive numerical simulations, we demonstrate that the 
normalized localization length $\beta$ of the eigenfunctions of the diluted non-Hermitian 
banded random matrix (nHdBRM) ensemble
scales with the parameter $x^*(N,b,\alpha)=\gamma(\alpha)[b_{\mbox{\tiny eff}}^2/N]^{\delta(\alpha)}$ 
as $\beta=x^*/(1+x^*)$; see Fig.~\ref{Fig01}(d).
Here, the effective bandwidth $b_{\mbox{\tiny eff}} \equiv \alpha b$ is the average number of non-zero 
elements per matrix row, $\alpha$ is the sparsity, $N$ is the matrix size, and $\gamma,\delta\sim 1$
are scaling parameters.
Moreover, we also verified that $x^*$ works well as the scaling parameter of the average inverse 
participation ratios $\left\langle \overline{\mbox{IPR}} \right\rangle$ (another eigenfunction measure) 
as well as the spectral properties of the nHdBRM ensemble, characterized by the ratio between nearest- 
and next-to-nearest-neighbor (complex) eigenvalue distances $\left\langle \overline{r}_\mathbb{C} \right\rangle$ 
and the ratio between consecutive singular-value spacings $\left\langle \overline{r}_{\tbox{SV}} \right\rangle$; 
see Fig.~\ref{Fig02}(d-f).
In addition, we also found that all these quantities ($\beta$, 
$\left\langle \overline{\mbox{IPR}} \right\rangle$, 
$\left\langle \overline{r}_\mathbb{C} \right\rangle$ and
$\left\langle \overline{r}_{\tbox{SV}} \right\rangle$)
are highly correlated; see Fig.~\ref{Fig03}.

In addition, for completeness, we also performed a detailed study of eigenfunction and spectral 
properties properties of the HdBRM ensemble~\cite{MFMR17b}; see the Appendix~\ref{app}.
Specifically, we report $\beta$, $\left\langle \overline{\mbox{IPR}} \right\rangle$, 
$\left\langle \overline{r}_\mathbb{C} \right\rangle$ and $\left\langle \overline{r}_\mathbb{R} \right\rangle$
where $r_{\mathbb{R}}$ is the ratio between consecutive eigenvalue spacings.
Indeed, for the HdBRM ensemble we made similar conclusions as for the nHdBRM ensemble:
That is, $\beta$ follows the scaling law of Eq.~(\ref{betax*}), see Fig.~\ref{Fig04}(d), and $x^*$ 
works well as the scaling parameter of $\left\langle \overline{\mbox{IPR}} \right\rangle$, 
$\left\langle \overline{r}_\mathbb{C} \right\rangle$ and $\left\langle \overline{r}_\mathbb{R} \right\rangle$,
see Fig.~\ref{Fig05}(d-f).
However, in contrast with the nHdBRM ensemble, for the HdBRM ensemble we found that
$\left\langle \overline{\mbox{IPR}} \right\rangle$ follows the same scaling law as $\beta$ does,
see Eq.~(\ref{IPRx*}) and Fig.~\ref{Fig05}(d).

Finally, we want to mention that both scalings~(\ref{betax*}) and~(\ref{IPRx*}) can be written in
the ``model independent" form (see e.g.~\cite{FM92,CGIFM92}):
\begin{equation}
\label{1/deff}
\frac{1}{d(N,b,\alpha)} \approx \frac{1}{d(\infty,b,\alpha)} + \frac{1}{d(N,N,1)} .
\end{equation}
Above, $d(N,b,\alpha)\equiv\exp[\bra S(N,b,\alpha) \ket]$ and $d(N,N,1)=\exp[S_{\tbox{RGE}}(N)]$ 
(the reference entropy) for scaling~(\ref{betax*}), while
$d(N,b,\alpha)\equiv \left\langle \mbox{IPR}(N,b,\alpha) \right\rangle$ and $
d(N,N,1)=\mathrm{IPR}_{\mathrm{GOE}}$ for scaling~(\ref{IPRx*}).
 
Since diluted RM models can be used as null models for random networks 
(i.e.~the adjacency matrices of complex networks are, in general, diluted random 
matrices), we believe that the nHdBRM ensemble may be used to model the adjacency matrices
of certain types of directed random networks; that is, those having banded adjacency matrices.
%Comment about the non-Hermitian Lloyd model and non-Hermitian disordered wires.
We hope our results may motivate a theoretical approach to the nHdBRM ensemble.

\section*{Acknowledgements}

J.A.M.-B. thanks support from CONACyT-Fronteras (Grant No.~425854) 
and VIEP-BUAP (Grant No.~100405811-VIEP2024), Mexico.

\appendix
\renewcommand{\thefigure}{A\arabic{figure}}
\setcounter{figure}{0}

\section{Scaling properties of Hermitian diluted banded random matrices}
\label{app}

\begin{figure*}[ht]
\centering
\includegraphics[width=0.65\textwidth]{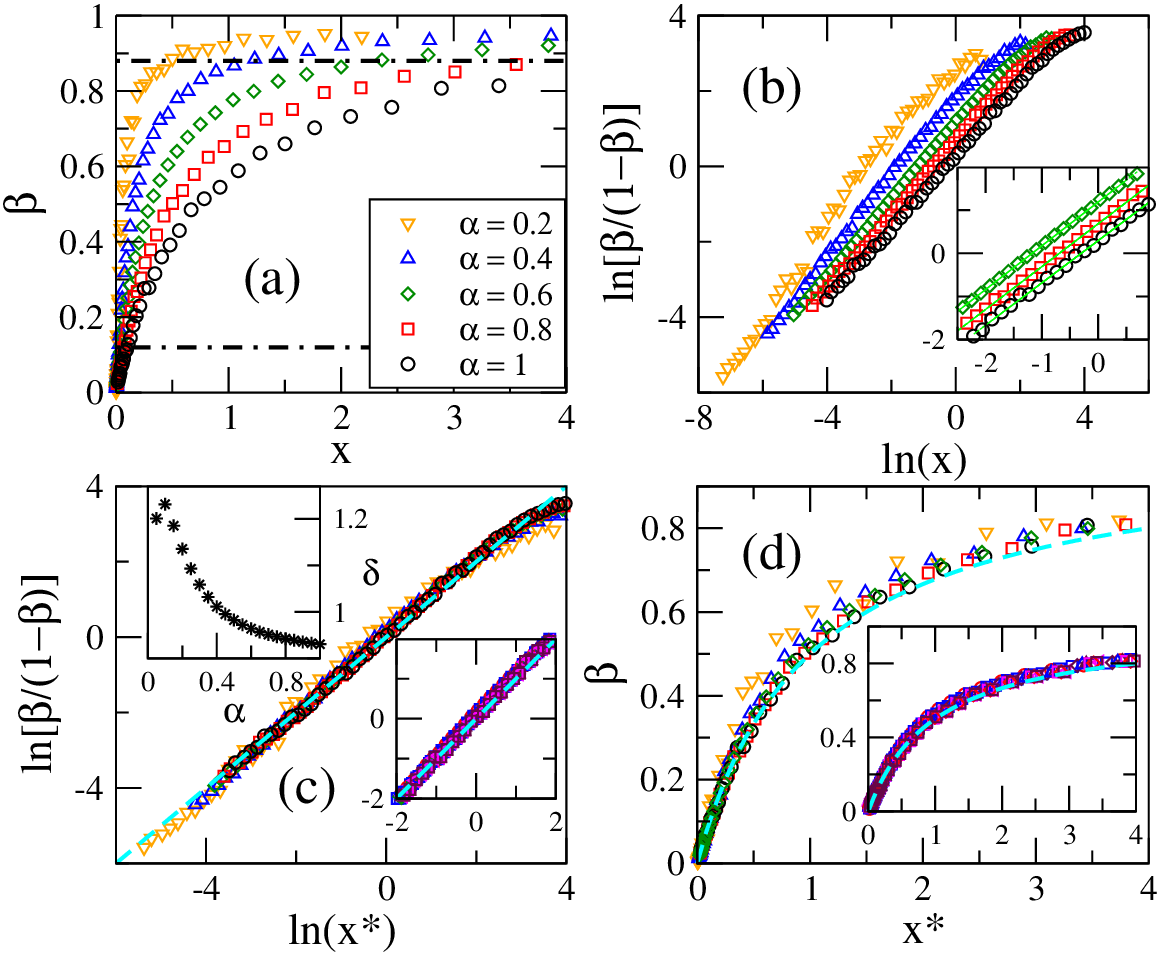}
\caption{
(a) Scaled localization length of eigenfunctions $\beta$ as a function of $x=b_{\mbox{\tiny eff}}^2/N$, 
$b_{\mbox{\tiny eff}}=\alpha b$, for the Hermitian diluted banded random matrix ensemble 
characterized by the sparsity $\alpha$. Several combinations of $(b,N)$ are used.
Horizontal dot-dashed lines at $\beta\approx 0.12$ and 0.88 are shown as a reference.
(b) Logarithm of $\beta/(1-\beta)$ as a function of $\ln(x)$. 
Inset: Enlargement in the range $\ln[\beta/(1-\beta)]=[-2,2]$ including data for 
$\alpha=0.6$, 0.8, and 1. Green-dashed lines are fittings of the data with Eq.~(\ref{betascaling2}).
(c) Logarithm of $\beta/(1-\beta)$ as a function of $\ln(x^*)$ [see Eq.~(\ref{betascaling3})]. 
Upper inset: Power $\delta$, from the fittings of the curves $\ln[\beta/(1-\beta)]$ vs.~$\ln(x)$ 
in the range $\ln[\beta/(1-\beta)]=[-2,2]$ with Eq.~(\ref{betascaling2}), as a function of $\alpha$. 
Lower Inset: Enlargement in the range $\ln[\beta/(1-\beta)]=[-2,2]$ including curves for 
$\alpha\in [0.5,1]$ in steps of $0.05$. Cyan-dashed lines in main panel and lower inset 
are Eq.~(\ref{betascaling3}).
(d) $\beta$ as a function of $x^*$. 
Inset: Data for $\alpha\in [0.5,1]$ in steps of $0.05$. Cyan-dashed lines in main 
panel and inset are Eq.~(\ref{betax*}).
}
\label{Fig04}
\end{figure*}
\begin{figure*}[ht]
\centering
\includegraphics[width=0.6\textwidth]{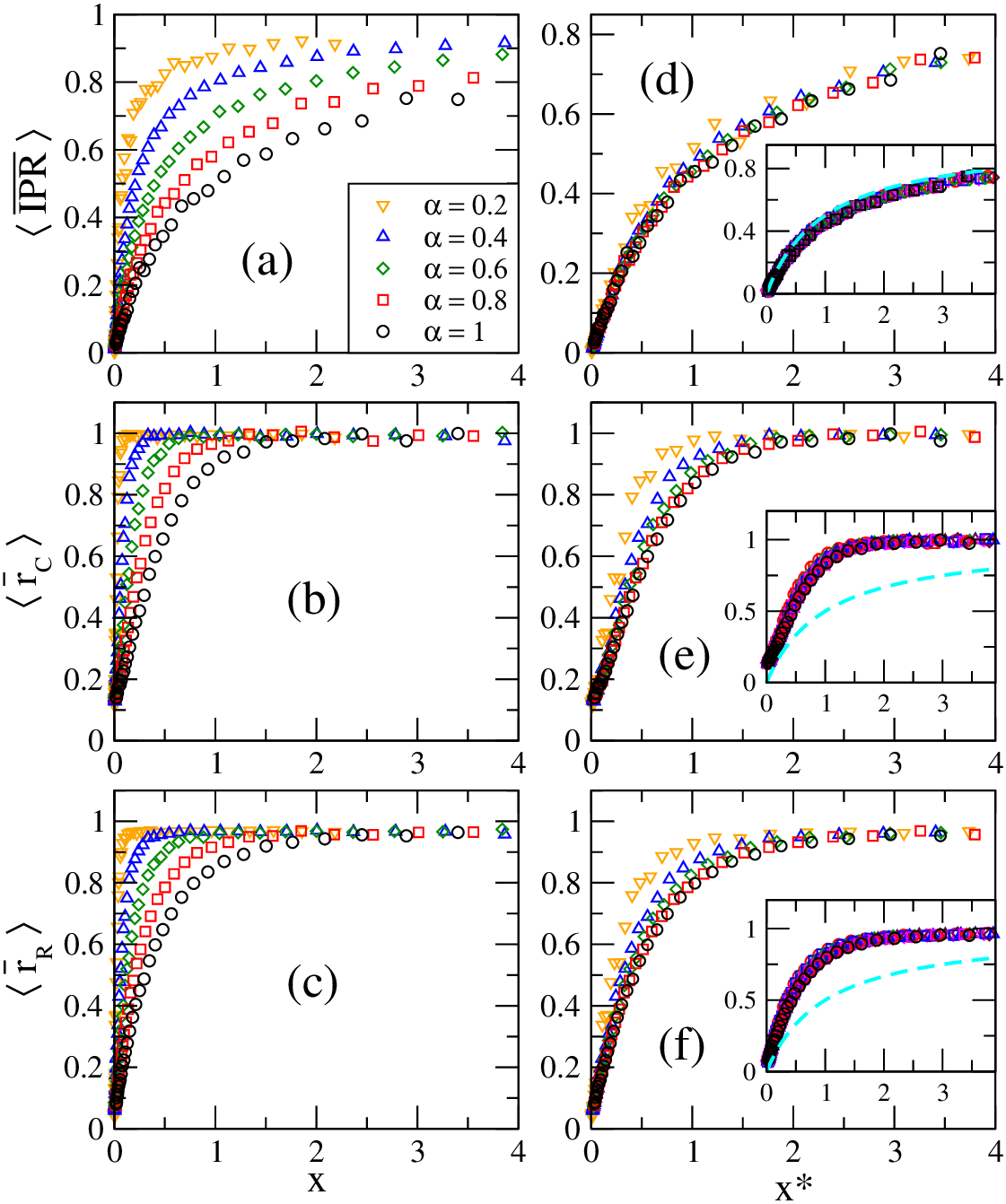}
\caption{
(a) $\left\langle \overline{\mbox{IPR}} \right\rangle$, 
(b) $\left\langle \overline{r}_\mathbb{C} \right\rangle$ and
(c) $\left\langle \overline{r}_\mathbb{R} \right\rangle$
as a function of $x$ for the Hermitian diluted banded random matrix ensemble characterized by the sparsity $\alpha$. 
(d) $\left\langle \overline{\mbox{IPR}} \right\rangle$, 
(e) $\left\langle \overline{r}_\mathbb{C} \right\rangle$ and
(f) $\left\langle \overline{r}_\mathbb{R} \right\rangle$
as a function of $x^*$.
Insets in (d-f): 
Data for $\alpha\in [0.5,1]$ in steps of $0.05$. Dashed lines are Eq.~(\ref{betax*}).
}
\label{Fig05}
\end{figure*}
\begin{figure*}[ht]
\centering
\includegraphics[width=0.65\textwidth]{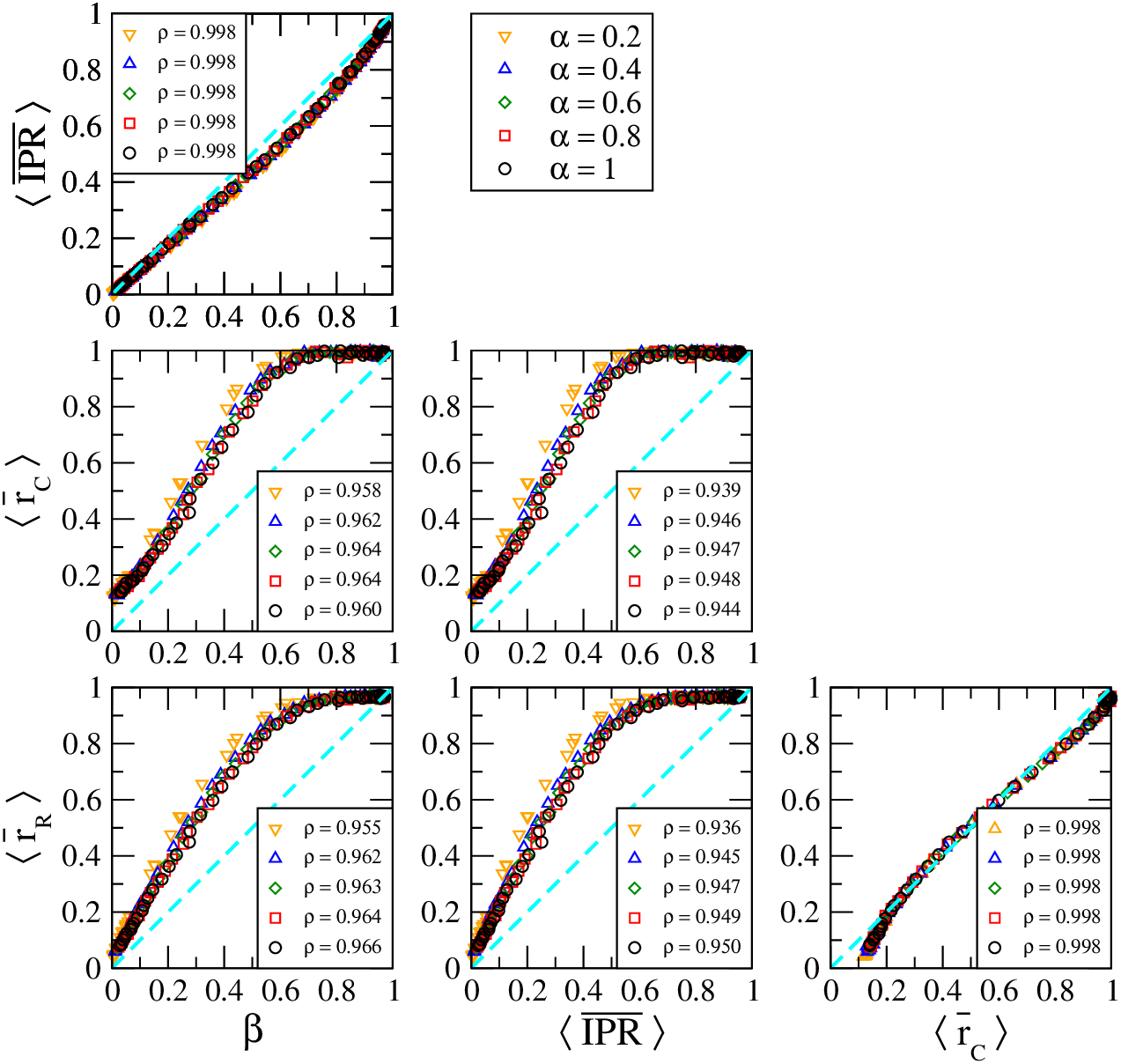}
\caption{
Scatter plots between $\beta$, $\left\langle \overline{\mbox{IPR}} \right\rangle$, 
$\left\langle \overline{r}_\mathbb{C} \right\rangle$ and
$\left\langle \overline{r}_\mathbb{R} \right\rangle$ for the Hermitian diluted banded 
random matrix ensemble. 
Pearson's correlation coefficients $\rho$ are reported in the corresponding panels.
}
\label{Fig06}
\end{figure*}

Here, we report the scaling of eigenfunction and spectral properties of the matrices ${\bf H}$ 
belonging to the HdBRM ensemble. This is done for two main reasons: 
First, for comparison purposes; so we can contrast eigenfunction and spectral properties of 
the HdBRM and the nHdBRM ensembles.
Second, for completeness; because in the study of the HdBRM ensemble presented in 
Ref.~\cite{MFMR17b} the spectral properties were only characterized by the repulsion parameter 
of Izrailev's distribution of eigenvalue spacings and the $\mathrm{IPR}$ of the eigenfunctions
was not reported.

Thus, as for the nHdBRM, we use exact numerical diagonalization to obtain the eigenfunctions 
$\Psi^m$ ($m=1\ldots N$) and eigenvalues of large ensembles of matrices ${\bf H}$ 
(which are members of the HdBRM ensemble) characterized by the parameters $N$, $b$, and $\alpha$. 
For each of the averages reported below we used at least $5\times 10^5$ data values. 

In Fig.~\ref{Fig04} we show the scaled localization length $\beta$ of eigenfunctions for the
HdBRM ensemble. We compute $\beta$ as in Eq.~(\ref{betaS}) with
\[
\ell_N = N \exp\left[ -\left( S_{\tbox{GOE}} - \bra S \ket \right)\right] \ ,
\]
where $S_{\tbox{GOE}}\approx\ln(N/2.07)$. This because
when $\alpha=1$ and $b=N$ the HdBRM becomes the GOE.
Notice that, for comparison purposes, Fig.~\ref{Fig04} is equivalent to Fig.~\ref{Fig01} 
for the nHdBRM ensemble. We also note that all the information in Fig.~\ref{Fig04} was reported in Figs.~1 
and~2 of Ref.~\cite{MFMR17b}, however, for completeness, we decided to include it here.
From Fig.~\ref{Fig04} we can validate the following conclusion made in Ref.~\cite{MFMR17b}:
The normalized localization length $\beta$ of the eigenfunctions of the HdBRM 
ensemble scales with the parameter $x^*(N,b,\alpha)=\gamma(\alpha)[(b\alpha)^2/N]^{\delta(\alpha)}$ 
as $\beta = x^*/(1+x^*)$, see Fig.~\ref{Fig04}(d), where $\gamma,\delta\sim 1$.

Then, in Fig.~\ref{Fig05} we report additional RMT measures to characterize eigenfunction and spectral 
properties of the HdBRM ensemble:
\begin{equation}
\langle \overline{\mathrm{IPR}}\rangle= \frac{\langle \mathrm{IPR} \rangle-\mathrm{IPR}_{\mathrm{PE}}}{\mathrm{IPR}_{\mathrm{GOE}}-\mathrm{IPR}_{\mathrm{PE}}} ,
\label{anIPRH}
\end{equation}
\begin{equation}
\langle \overline{r}_\mathbb{C}\rangle= \frac{\langle r_\mathbb{C}  \rangle-r_{\mathbb{C}_{\mathrm{PE}}}}{r_{\mathbb{C}_{\mathrm{GOE}}}-r_{\mathbb{C}_{\mathrm{PE}}}} 
\label{anrCH}
\end{equation}
and
\begin{equation}
\langle \overline{r}_\mathbb{R}\rangle= \frac{\langle r_\mathbb{R}  \rangle-r_{\mathbb{R}_{\mathrm{PE}}}}{r_{\mathbb{R}_{\mathrm{GOE}}}-r_{\mathbb{R}_{\mathrm{PE}}}} ,
\label{anrR}
\end{equation}
The reference values used in Eqs.~(\ref{anIPRH}-\ref{anrR}), corresponding to the GOE, are reported in 
Table~\ref{T1}.
Note that Figs.~\ref{Fig05}(a,b,d,e) are equivalent to Figs.~\ref{Fig02}(a,b,d,e) for the nHdBRM ensemble.
Also notice that here, for the HdBRM ensemble, we are not computing $\langle \overline{r}_{\tbox{SV}}\rangle$
since the spectrum of ${\bf H}$ is real and we do not need SVS. Instead, given the real and ordered spectrum 
of the Hermitian matrices ${\bf H}$, $\lambda_1>\lambda_2>\cdots >\lambda_n$, we compute the ratio 
$r_{\mathbb{R}}$ between consecutive level spacings, where the $i$--th ratio is given by~\cite{ABGR13}
\begin{equation}
r_{\mathbb{R}}^i = \frac{\mathrm{min}(\lambda_{i+1}-\lambda_i,\lambda_{i}-\lambda_{i-1})}{\mathrm{max}(\lambda_{i+1}-\lambda_i,\lambda_i-\lambda_{i-1})} .
\label{ratioreal}
\end{equation}

From Fig.~\ref{Fig05} we observe that $x^*$ works well as scaling parameter of 
$\left\langle \overline{\mbox{IPR}} \right\rangle$, 
$\left\langle \overline{r}_\mathbb{C} \right\rangle$ and
$\left\langle \overline{r}_\mathbb{R} \right\rangle$ for the HdBRM ensemble, see Figs.~\ref{Fig05}(d-f).
Perticularly for $\alpha\ge 0.5$, where all curves $\left\langle \overline{X} \right\rangle$ vs.~$x^*$ fall 
one on top of the other; see the corresponding insets.
Here, $X$ represents $\mbox{IPR}$, $r_\mathbb{C}$ and $r_\mathbb{R}$.
Moreover, remarkably, $\left\langle \overline{\mbox{IPR}} \right\rangle$ approximately follows the same
scaling law as $\beta$:
\begin{equation}
\left\langle \overline{\mbox{IPR}} \right\rangle \approx \frac{x^*}{1+x^*} ,
\label{IPRx*}
\end{equation}
see the inset in Fig.~\ref{Fig05}(d). 

Finally, in Fig.~\ref{Fig06} we present the scatter plots between $\beta$, 
$\left\langle \overline{\mbox{IPR}} \right\rangle$, 
$\left\langle \overline{r}_\mathbb{C} \right\rangle$ and
$\left\langle \overline{r}_\mathbb{R} \right\rangle$ for the HdBRM ensemble. 
We also include the Pearson's correlation coefficients $\rho$ in the corresponding panels.
Remarkably, we observe better correlations between eigenfunction measures 
(i.e.~$\left\langle \overline{\mbox{IPR}} \right\rangle$ vs.~$\beta$) and spectral measures
(i.e.~$\left\langle \overline{r}_\mathbb{R} \right\rangle$ vs.~$\left\langle \overline{r}_\mathbb{C} \right\rangle$)
for the HdBRM ensemble (where $\rho$ are very close to one) as compared with the nHdBRM ensemble.

\bibliographystyle{plainnat}

\end{document}